\documentclass[useAMS,usenatbib]{mn2e}
\usepackage{graphicx}
\usepackage{epstopdf}
\usepackage{amsmath}
\usepackage{amssymb}
\title[Thermally driven wind]{Thermal wind from hot accretion flows at large radii}
\author[Bu \& Yang]{De-Fu Bu$^1$\thanks{E-mail: dfbu@shao.ac.cn (DB)} and Xiao-Hong Yang$^2$\thanks{E-mail: yangxh@cqu.edu.cn} \\
$^{1}$Key Laboratory for Research in Galaxies and Cosmology, Shanghai Astronomical Observatory, Chinese Academy of Sciences,\\ 80 Nandan Road, Shanghai 200030, China \\
$^2$ Department of Physics, Chongqing University, Chongqing 400044, China}
\begin{document}

%\date{Accepted 1988 December 15. Received 1988 December 14; in original form 1988 October 11}

\pagerange{\pageref{firstpage}--\pageref{lastpage}} \pubyear{2002}

\maketitle

\label{firstpage}
\begin{abstract}
 We study slowly rotating accretion flow at parsec and sub-parsec scale irradiated by a low luminosity active galactic nuclei. We take into account the Compton heating, photoionization heating by the central X-rays. The bremsstrahlung cooling, recombination and line cooling are also included. We find that due to the Compton heating, wind can be thermally driven. The power of wind is in the range $(10^{-6}-10^{-3}) L_{\rm Edd}$, with $L_{\rm Edd}$ being the Eddington luminosity. The mass flux of wind is in the range $(0.01-1) \dot M_{\rm Edd}$ ($\dot M_{\rm Edd}= L_{\rm Edd}/0.1c^2$ is the Eddington accretion rate, $c$ is speed of light). We define the wind generation efficiency as $\epsilon = P_W/\dot {M}_{\rm BH}c^2$, with $P_W$ being wind power, $\dot M_{\rm BH}$ being the mass accretion rate onto the black hole. $\epsilon$ lies in the rage $10^{-4}-1.18$. Wind production efficiency decreases with increasing mass accretion rate. The possible role of the thermally driven wind in the active galactic feedback is briefly discussed.
\end{abstract}

\section{Introduction}
Hot accretion flow is believed to operate in low-luminosity active
galactic nuclei (LLAGNs; e.g. Ho 2008; Antonucci 2012; Done 2014)
and the hard/quiescent states of black hole X-ray binaries (e.g.
Esin et al. 1997; Fender et al. 2004; Zdziarski \& Gierli\'{n}ski
2004; Remillard \& McClintock 2006; Narayan \& McClintock 2008; Bolloni
2010; Wu et al. 2013; Yuan \& Narayan 2014). Hot accretion flow was
studied analytically in the 1990s (Narayan \& Yi 1994, 1995;
Abramowicz et al. 1995; Kato et al. 1998, Narayan et al. 1998).
Later on, numerical simulations have been performed to study the
properties of hot accretion flow (e.g. Stone et al. 1999;
Igumenshchev \& Abramowicz 1999; 2000; Hawley et al. 2001; Machida
et al. 2001; De Villiers et al. 2003; Pen et al. 2003; Beckwith et
al. 2008; Pang et al. 2011; Tchekhovskoy et al. 2011; McKinney et
al. 2012; Yuan et al. 2012a; 2012b; Sadowski et al. 2013;
Mo\'{s}cibrodzka et al. 2014).

One of the most important results found by numerical simulations is
that strong wind exists in hot accretion flow (e.g. Yuan et al.
2012b; Yuan et al. 2015; Narayan et al. 2012; Li et al. 2013; see also Moller \& Sadowski
2015). %In
%the hydrodynamic (HD) simulation, wind is found to be driven by the
%combination of gas pressure gradient and centrifugal forces. In the
%magneto-hydrodynamic (MHD) simulation, wind is found to be driven by
%the combination of gas pressure and magnetic pressure gradients and
%centrifugal forces (Yuan et al. 2015; see also Moller \& Sadowski
%2015).
Recently, observations of both LLAGNs (e.g. Crenshaw \&
Kramemer 2012; Tombesi et al. 2010; 2014; Wang et al. 2013; Cheung
et al. 2016) and the hard state of black hole X-ray binaries (Homan
et al. 2016) show that winds are present in hot accretion flow.
Therefore, the numerical simulation result is confirmed by
observations.

The studies mentioned above focus on the flow at the region very
close to the black hole. The outer boundary of the simulations is
several hundreds of $r_s$ ($r_s$ is Schwarzschild radius). Then a
question is that what are the properties of accretion flow beyond
hundreds of $r_s$? Can the result from the above mentioned
simulations be applied to the accretion flow at sub-parsec and
parsec scale? The sub-parsec and parsec scale accretion flow
connects the flow in the AGNs and the flow beyond the Bondi radius.
The feeding gas of the central black hole comes from this region.
Therefore, it is important to study the properties of accretion flow
at sub-parsec and parsec scales.

Recently, Li et al. (2013) and Inayoshi et al. (2017) study the
accretion flow at parsec and sub-parsec scales. They find that if the gas
density is low enough and radiation is not important, the properties
of the flow at parsec and sub-parsec scale are similar to the flow inside hundreds of $r_s$. We
note that in these two works, the AGNs feedback is not
taken into account. However, AGNs feedback effects (e.g. momentum feedback and energy feedback) are very
important to determine the properties of the flow at parsec scale
(e.g. Magorrian et al. 1998;
Ferrarese \& Merritt 2000; Gebhardt et al. 2000;  Di Matteo et al.
2005; Ciotti \& Ostriker 2007; Kormendy \& Bender 2009; Ostriker et al. 2010;
Yuan \& Li 2011; Liu et al. 2013; Gan et al. 2014)
and need to be taken into account.

Recently, the accretion flow at sub-parsec and parsec scale
irradiated by a quasar is studied (e.g. Proga 2007; Kurosawa \& Proga 2009). It is found that outflows at
parsec-scale can be driven by line force due to interaction of UV
photons and the not fully ionized gas.

In this paper, we focus on the accretion flow at sub-parsec and
parsec scale irradiated by a LLAGN. Different from a quasar, LLAGNs
emit the majority of photons in X-ray band (Yuan \& Narayan 2014).
The X-ray from the vicinity of black hole can heat the gas at
sub-parsec and parsec scale. If the radiation is strong enough, the
temperature of the gas at parsec scale will be heated to be above
the virial temperature, so that the accretion process will be
stopped. In this paper, in addition to the Compton heating/cooling,
we also consider the photoionization heating-recombination cooling,
bremsstrahlung cooling and line cooling. In this paper, as a first
step, we consider gas with small angular momentum.

The paper is organized as follows. In section 2, we describe our
models and method; In section 3, we present our results; Section 4
is devoted to conclusions and discussions.

\section{Numerical method and models }
In this paper, we study the accretion flow at sub-parsec and parsec
scales. \emph{We define LLAGN to be the accretion flow inside the inner
boundary of our simulation which is not resolved.} We
assume that all the photons emitted by the central LLAGN are in X-ray
band. The luminosity of the LLAGN is self-consistently determined
based on the mass accretion rate through the inner boundary. We set
that the inner boundary of the computation domain is much larger
than the radius ($20r_{\text{s}}$) inside which most of the
radiation is produced by the LLAGN. Therefore, the radiation of LLAGN
can be approximated to be from a point object located at $r=0$ and
be isotropic.

In this work, we take 2$\% L_{\rm Edd}$ to be the upper limit of the
luminosity of the central LLAGN. Above 2$\% L_{\rm Edd}$, the spectrum of a black hole may
transit from hard to soft state (e.g. Yuan \& li 2011). We only consider the cases in which the luminosity of central LLAGN is below 2$\% L_{\rm Edd}$.

\subsection{Basic equations}
We assume the accretion flow to be axisymmetric. We use the ZEUS-MP code (Hayes et al. 2006) in spherical coordinates ($r,\theta,\phi$)
to solve the HD equations below:

\begin{equation}
 \frac{d\rho}{dt} + \rho \nabla \cdot {\bf v} = 0,
\end{equation}
\begin{equation}
 \rho \frac{d{\bf v}}{dt} = -\nabla p - \rho \nabla \Phi %+ \rho \bf{F_{rad}}
\end{equation}
\begin{equation}
 \rho \frac{d(e/\rho)}{dt} = -p\nabla \cdot {\bf v} + \dot E
\end{equation}
Here, $\rho$ is the mass density, $\bf v$ is the velocity, $p$ is
gas pressure, $e$  is internal energy. We adopt an equation of state of ideal gas $p=(\gamma-1)e$ with adiabatic index $\gamma=5/3$. $\Phi=-GM/(r-r_s)$ is the gravitational potential, where $M$ and $G$ are the central black hole mass and the gravitational constant, respectively. The Schwarzschild radius $r_s=2GM/c^2$.

Because the luminosity of the central LLAGN is below 2$\% L_{Edd}$, the radiation pressure due to Compton scattering is not important. Therefore, we do not take into account the radiation pressure force in the momentum Equation (2).

In Equation (3), $\dot{E}$ is the net gas energy change rate due to
heating and cooling. We consider Compton heating/cooling,
bremsstrahlung cooling, photoionization heating, and line and
recombination cooling (Sazonov et al. 2005).
\begin{equation}
\dot E = n^2(S_C+S_{br}+S_{ph,rec,l})
\label{heating}
\end{equation}
In this equation, $n$ is gas number density. The Compton heating/cooling is:
\begin{equation}
S_C=4.1 \times 10^{-35} (T_X-T) \xi     %{\rm erg cm^3 s^{-1}}
\label{Compton}
\end{equation}
In Equation (\ref{Compton}), $T$ is gas temperature. $T_X$ is the radiation temperature determined by the spectrum of
radiation from the central source. For a quasar, $T_X=1.9\times10^7$ K (Sazonov et al.(2005). For a
LLAGN, $T_X$ will be much higher (Yuan et al.
2009; Xie et al. 2017). Here, we set $T_X=10^8$ and $10^9
\text{K}$ for the comparison of effect of Compton temperature on
results. $\xi$ is the photoionization parameter
defined by
\begin{equation}
\xi\equiv\frac{L}{nr^2}e^{-\tau_{\text{x}}},
\end{equation}
where $\tau_{\text{x}}$ ($=\int_0^r \rho \kappa_{\text{x}}dr$, where
$\kappa_{\text{x}}$ is the X-ray opacity) is the X-ray scattering
optical depth in the radial direction, $r$ is the distance from the
central source, the number density of local gas $n=\rho/(\mu m_{\text{p}})$, $\mu$ is the mean molecular
weight. We set $\mu=1$ and $\kappa_{\text{x}}=0.4
\text{cm}^2\text{g}^{-1}$ because Thomson scattering dominates the
attenuation.
The bremsstrahlung cooling can be expressed as
\begin{equation}
S_{br}=-3.8 \times 10^{-27} \sqrt{T}     %{\rm erg cm^3 s^{-1}}
\label{brem}
\end{equation}

We do not show the exact formula of the sum of photoionization heating, line and recombination cooling. The reason is as follows. Sazonov et al. (2005) shows that when $T> 10^7$ K, $S_{ph,rec,l} \ll S_C, S_{br}$.
In this paper, we find that the gas temperature is always comparable to or higher than $10^7$ K. Therefore, the dominant heating and cooling processes are Compton heating and bremsstrahlung cooling. For the formula of $S_{ph,rec,l}$, we refer to Equation (A35) of Sazonov et al. (2005).

In reality, the emitted photons by the accretion flow can interact with the accretion flow and change the properties of the flow. In this paper, we do not consider the secondary heating/cooling by the radiation emitted by the gas.

\subsection{Luminosity of the central LLAGNs}
Luminosity of the central LLAGN is determined by the accretion rate
onto the black hole and radiative efficiency. The radiative
efficiency depends on the accretion rate and the
parameter $\delta$ which describes the fraction of the direct
viscous heating to electrons. %Xie \& Yuan (2012) It is showed that
%radiative efficiency of hot accretion flow increases with the
%accretion rate and $\delta$.
%According to the results of Xie \& Yuan (2012), we couple the
%luminosity of the central LLAGN and the inflow mass rate through the inner boundary of
%computational domain in the following way.

We define the `circularization' radius ($r_{\text{cir}}$) to be the
radius at which the Keplerian angular momentum is equal to the
specific angular momentum of the accretion gas. We study low angular momentum accretion flow by setting
$r_{\text{cir}}$ to be smaller than the inner boundary of the
computational domain. It means that the angular momentum of gas is much smaller than the Keplerian angular momentum of the computational domain. In this case, the shear of rotational velocity in radial direction is negligibly small. Viscosity is proportional to shear of rotational velocity (see Stone, Pringle \& Begelman 1999). Therefore, viscosity is also negligibly small. The negligibly small viscosity can not re-distribute angular momentum of gas. Therefore, when the gas falls from large to small radii, the angular momentum is conserved and does not depends on radius. In previous work (Bu et al. 2013), we find that outside of $r_{\text{cir}}$, there is no outflow and the accretion rate is a constant with radius. Thus, we assume that the mass
accretion rate is constant with radius between the inner boundary of the computational domain
and $r_{\text{cir}}$. Gas can fall onto the black hole from $r_{\text{cir}}$ in the presence of viscosity. Previous works show that the mass inflow rate of  hot
accretion flow with large angular momentum can be described as
$\dot{M}_{\text{in}}\propto r^s$, where $s\approx0$ for
$r\lesssim 10 r_{\text{s}}$ and $s\approx0.5$ for
$10r_s<r<r_{\text{cir}}$ (Yuan et al. 2012, 2015; Bu et al. 2013,
2016a, 2016b). The accretion rate of the central black hole is calculated as,
\begin{equation}
\dot{M}_{\text{BH}}=\dot{M}_{\text{in}}
(\frac{10r_{\text{s}}}{r_{\text{cir}}})^{0.5},
\label{AccretionrateBH}
\end{equation}
In equation (\ref{AccretionrateBH}), $\dot{M}_{\text{in}}$ is the mass inflow rate at the
inner boundary of computational domain. According to equation (\ref{AccretionrateBH}), the mass inflow rate is constant with radius between $r_{\text{cir}}$ and the inner boundary of simulation domain. From $r_{\text{cir}}$ to $10r_{\text{s}}$, the mass inflow rate decreases inwards. From $10r_{\text{s}}$ to the black hole horizon, mass inflow rate is constant with radius. After gas falls into the inner boundary of the simulation, one portion of gas will be accreted to the black hole, the other portion of gas will form wind and move outward to larger radii. Therefore, density of gas between black hole and the inner boundary of simulation does not keep increasing with time. In this paper, as a first step, we neglect the mechanical feedback effects by wind generated between 10$r_{\text{s}}$ and $r_{\text{cir}}$. The radiative efficiency calculated by Xie \& Yuan (2012)
in the case of viscous parameter $\alpha=0.1$ can be described as follows,
\begin{equation}
\epsilon(\dot{M}_{\text{BH}})=\epsilon_0(\frac{100\dot{M}_{\text{BH}}}{\dot{M}_{\text{Edd}}})^a,
\end{equation}
where $\epsilon_0$ and $a$ are given in Table 1 of Xie \& Yuan (2012) for different $\delta$. We choose the case of $\delta=0.5$ and have
\begin{equation}
(\epsilon_0,a) = \left\{ \begin{array}{ll}
(1.58,0.65) & \textrm{if } \frac{\dot{M}_{\text{BH}}}{\dot{M}_{\text{Edd}}}\lesssim2.9\times10^{-5};\\
(0.055,0.076) & \textrm{if } 2.9\times10^{-5}<\frac{\dot{M}_{\text{BH}}}{\dot{M}_{\text{Edd}}}\lesssim3.3\times10^{-3};\\
(0.17,1.12) & \textrm{if } 3.3\times10^{-3}<
\frac{\dot{M}_{\text{BH}}}{\dot{M}_{\text{Edd}}}\lesssim5.3\times10^{-3}.
\end{array} \right.
\end{equation}
When $\frac{\dot{M}_{\text{BH}}}{\dot{M}_{\text{Edd}}}>5.3\times10^{-3}$,
the radiative efficiency $\epsilon(\dot{M}_{\text{BH}})$ is simply
set to be 0.1.

Time is needed for gas moving from the inner
boundary of computation domain to black hole. We need to consider the time lag between X-ray photons generation close to the black hole and the accretion rate calculated at the inner boundary of the computational domain (Kurosawa \& Proga 2009). Accretion luminosity at a given time is calculated by $L(t)=\epsilon
{\dot{M}}_{\text{BH}}(t-\delta t) c^2$.
${\dot{M}}_{\text{BH}}(t-\delta t)$ is the net mass
accretion rate calculated at time $t-\delta t$. $\delta t$ is the lag time and equaling to the accretion timescale from the inner boundary of the simulation domain to the black hole horizon. As introduced above, gas can freely fall from the inner boundary to $r_{\text{cir}}$. Therefore, the time needed for gas traveling from inner boundary of simulation to $r_{\text{cir}}$ is $\delta t_1=r_{in}/V1$, $V1$ is free fall velocity at $r_{in}$. The time needed for gas traveling from $r_{\text{cir}}$ to $10 r_s$ is the viscous timescale at $r_{\text{cir}}$ and is calculated by $\delta t_2=r_{\text{cir}}/V2/\alpha$. $V2$ is free fall velocity at $r_{\text{cir}}$, $\alpha$ is viscous coefficient. The time needed for gas traveling from $10r_s$ to black hole is $\delta t_3=10r_s/V3$, $V3$ is free fall velocity at $10 r_{s}$. The total time delay $\delta t=\delta t_1+\delta t_2+\delta t_3$. The delay time $\delta t$ is calculated based on physical considerations. In order to test the effects of changing value of $\delta t$, we have carried out model 1aLDT. In model 1aLDT, we set the delay time is 10 times that in model 1a. We find that with the increase of $\delta t$, the wind episode becomes slightly shorter and wind becomes slightly violent. The result of model 1aLDT is presented in Table 1.

We assume the black hole is located at $r=0$. The mass of the black
hole is set to be $M=10^8M_{\odot}$ ($M_{\odot}$ is solar mass). The
computational domain covers a range $500 r_{\text{s}} \leq r \leq
10^6 r_{\text{s}}$ ($5\times 10^{-3}$ parsec $\leq r \leq 10$ parsec) in radial direction and $0 \leq \theta \leq
\pi/2$ in $\theta$ direction. The radial grids are logarithmically
spaced ($dr_{i+1}/dr_i=1.05$). In $\theta$ direction, the grids are
uniformly spaced. The resolution in is paper is $140 \times 88$.
Axis-of-symmetry and reflecting boundary conditions are applied at
the pole (i.e. $\theta=0$) and the equatorial plane (i.e.
$\theta=\pi/2$), respectively. The outflow boundary condition is
adopted at the inner radial boundary. At the outer radial boundary ($r_{\text{out}}$),
all HD variables except the radial velocity are set to be equal to
the initially chosen values when $v_r(r_{\text{out}},\theta)<0$
(inflowing); all HD variables in the ghost zones are set to the
values at $r_{\text{out}}$, when $v_r(r_{\text{out}},\theta)>0$
(outflowing).

We assume the initial density and temperature of the accretion gas
are uniform with $\rho=\rho_0$ and $T=T_0$. We also assume that
initially $v_r=v_{\theta}=0$. The angular momentum of the initial
condition is set to be $l(\theta)=l_0 (1-|cos(\theta)|)$, where
$l_0$ equals to the Keplerian angular momentum at the equatorial
plane at $r_{\text{cir}}$.

\begin{table*} \caption{Simulation parameters }
\begin{tabular}{cccccccccc}
\hline \hline
 Model &  $T_{\text{X}}$ & $\rho_0$                    & $T_0$     & $\tau_{\text{x}}$ &  $\dot M_{\rm in}(r_{\rm in})$ & $\dot M_{\rm BH}$ & $\dot M_{\rm w}(r_{\rm out})$ & $P_W(r_{\rm out})$         & $\epsilon(r_{\rm out})$\\
  &  ($10^8$K)      & ($10^{-22}\text{g cm}^{-3}$)& ($10^7$K) & (at $R_{\text{c}})$ & ($L_{\rm Edd}/0.1c^2$)         & ($L_{\rm Edd}/0.1c^2$)    & ($L_{\rm Edd}/0.1c^2$)     & ($L_{\rm Edd}$) &    \\
(1) & (2)             & (3)                         &  (4)      &     (5)          &        (6)                     &     (7)          &   (8)           &     (9)         &    (10)\\

\hline\noalign{\smallskip}
1a   & 1 &1       &   2   & 0.015 & $10^{-2}$ & $1.7 \times 10^{-3}$& $6.5\times 10^{-2}$ &$1.7 \times 10^{-6}$ & $10^{-4}$\\
1aLDT  & 1 &1       &   2   & 0.019 & $1.3 \times 10^{-2}$ & $2.2 \times 10^{-3}$& $9.1\times 10^{-2}$ &$2.55 \times 10^{-6}$ & $1.16 \times 10^{-4}$\\
2a   & 1 &1       &   4   & 0.013 & $7 \times 10^{-4}$ & $1.2 \times 10^{-4}$ & $2\times 10^{-1}$ & $1.5 \times 10^{-5}$ & $1.25\times 10^{-2}$ \\
3a   & 1 &1       &   6   & $6 \times 10^{-4}$ & $2\times10^{-5}$ & $3.4\times10^{-6}$ & $1.1\times 10^{-1}$ & $1.7 \times 10^{-5}$ & $0.5$ \\
4a   & 1 &1       &   10  & $2.5 \times 10^{-4}$ & $10^{-5}$ & $1.7 \times 10^{-6}$ & $1\times 10^{-1}$ & $2.2 \times 10^{-5}$ & $1.18$ \\
%5a   & 1 &3       &   2   & 350 & $4\times 10^{-2}$ & $6.8 \times 10^{-3}$ & & $1.2 \times 10^{-5}$ & $1.7 \times 10^{-4}$ \\
2ad   & 1 &3       &   4   & 0.025 & $3\times 10^{-2}$ & $5.1\times 10^{-3}$& $6\times 10^{-1}$ &  $4 \times 10^{-5}$ & $8 \times 10^{-4}$\\
3ad   & 1 &3       &   6   & 0.018 & $1.2\times 10^{-2}$& $2\times 10^{-3}$  & $8\times 10^{-1}$ &$8\times 10^{-5}$&$4\times 10^{-3}$ \\
4ad   & 1 &3       &   10   & 0.021 & $1.7\times 10^{-3}$ & $2.9\times 10^{-4}$ & $9.2\times 10^{-1}$ & $1.8\times 10^{-4}$ & $6.2\times 10^{-2}$ \\
1aTx   & 10 &1       &   2   & 0.01 & $1.2\times 10^{-3}$ & $2\times 10^{-4}$  & $1.5\times 10^{-1}$  & $5.5\times 10^{-6}$  & $2.8\times 10^{-3}$ \\
2aTx   & 10 &1       &   4   & 0.009 & $3.6\times 10^{-4}$ & $6.1\times 10^{-5}$ & $2\times 10^{-1}$  &$1.6\times 10^{-5}$  & $2.6\times 10^{-2}$  \\
3aTx   & 10 &1       &   6   & $2 \times 10^{-4}$ & $2.2\times 10^{-5}$ & $3.7\times 10^{-6}$ & $1.1\times 10^{-1}$ & $1.4\times 10^{-5}$ & $0.38$\\

 \hline\noalign{\smallskip}
\end{tabular}

Note: Col. 1: model names. Col.2: the Compton temperature of X-ray radiation from central AGNs powered by a hot accretion flow. Cols 3, 4: the density and temperature at the outer bounary, respectively. Col 5: X-ray optical depth measured at the Compton radius. The optical depth is obtained by integration from inner boundary of simulation to Compton radius. Col. 6: the mass accretion rate measured at the inner boundary of the simulation domain. Col. 7: the mass accretion rate onto the central black hole (see Equation (\ref{AccretionrateBH}) for reference). Col. 8: the mass flux of wind measured at the outer boundary. Col. 9: the power carried by wind measured at the outer boundary (see Equation (\ref{power}) for definition). Col. 10: the production efficiency of wind (see Equation (\ref{efficiency}) for definition). Note that the results listed in Cols 5-10 are obtained by time average the data from $t=1.5\times 10^5$ to $10^6$ year.

\end{table*}

\section{Results}

We summarize all the models with different parameters in Table 1. We examine the effects of changing density ($\rho_0$) and temperature ($T_0$) at the outer boundary. We also examine the effects of changing the radiation temperature $T_X$.
Virial temperature at the outer boundary equals $\sim 10^7
\text{K}$. We consider temperature at the boundary is higher than
$T_0=10^7$K. Therefore, Bondi radius is located within the computational
domain and much less than the outer boundary.

We take model 1a as our fiducial model. In this model, we set
$\rho_0=10^{-22} \rm {g \cdot cm^{-3}}$ and $T_0=2\times 10^7 \rm K$. According
to the temperature, the Bondi radius in this model is
located at $1.6\times 10^5 r_s$. The Compton radius, i.e., the
radius at which the local isothermal sound speed (at the Compton
temperature, $T_X$) is equal to the escape velocity, is defined as
\begin{equation}
R_c=\frac{GM \mu m_{\rm H}}{k T_X}
\end{equation}
where $\mu=1$. Given that $T_X=10^8K$,
we have $R_c=9\times 10^4 r_s$. We run the simulation from $t=0$
to $t=1.15\times 10^6 $ year. We find that wind is generated episodically. The logic is as follows. When the
gas accretes to the black hole, X-ray photons will be generated and
the accretion flow will be heated by the X-ray irradiation. When the gas
temperature is increased to be above the Virial value, wind will
be formed. Wind takes away the fueling gas of the black hole, accretion rate onto the black hole will be suppressed. Then, the X-ray generation by the central LLAGN is suppressed. The gas at
sub-parsec and parsec scale will cool down when the X-ray heating is
suppressed and fuel to the central black hole again. Then the wind
can be generated again by the X-ray heating of the infalling gas.

\begin{figure*}
\begin{center}
\includegraphics[scale=0.38]{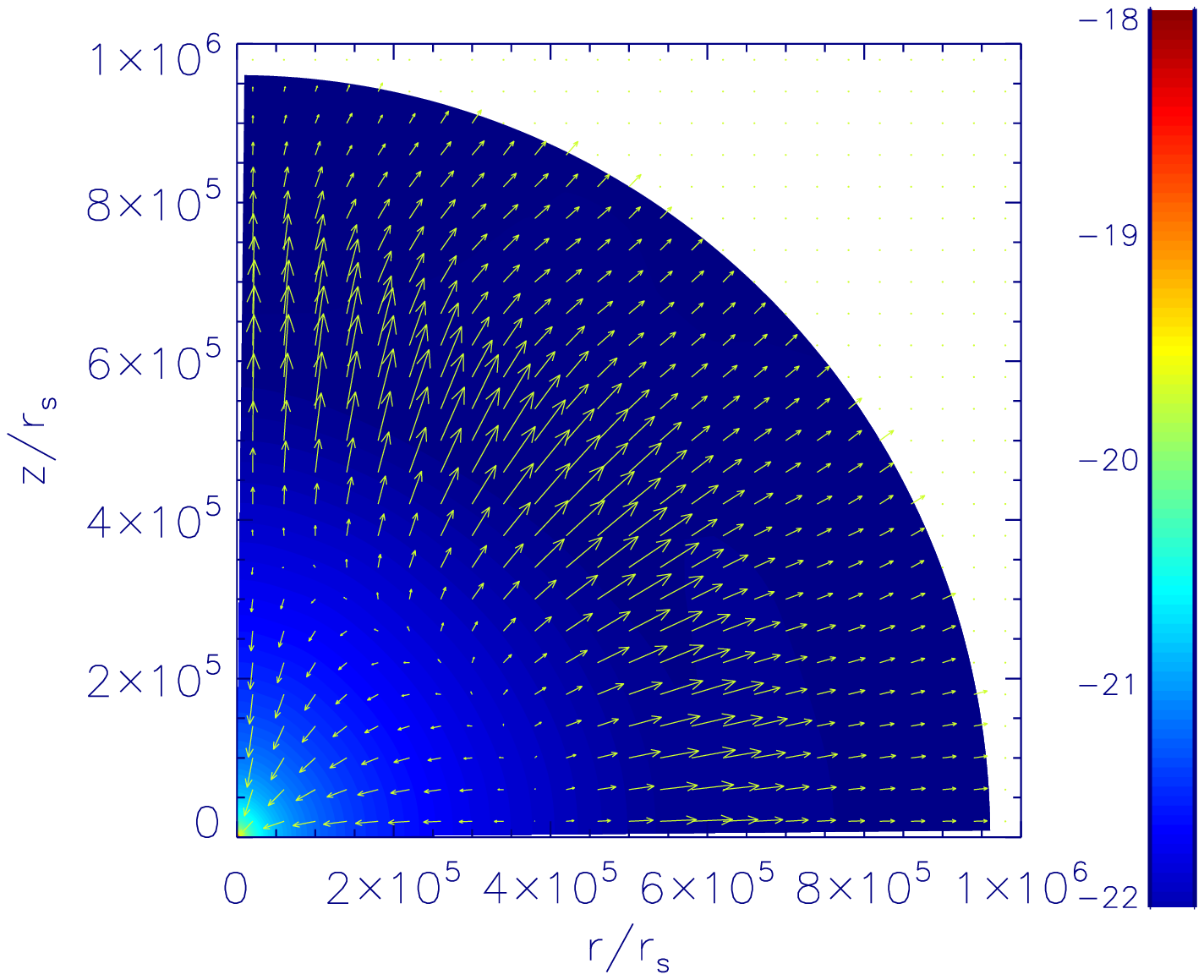}\hspace*{0.5cm}
\includegraphics[scale=0.38]{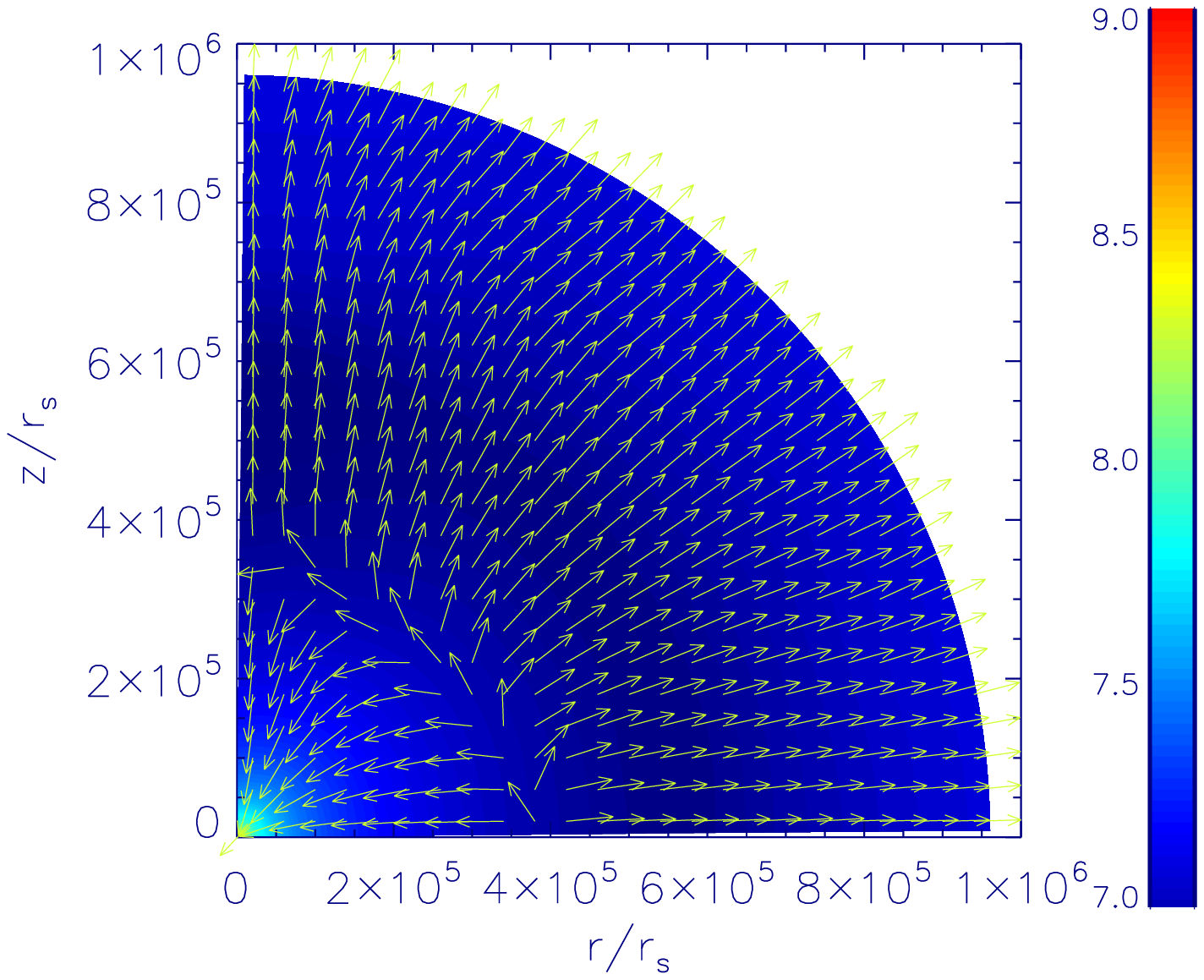}\hspace*{0.5cm}
\includegraphics[scale=0.38]{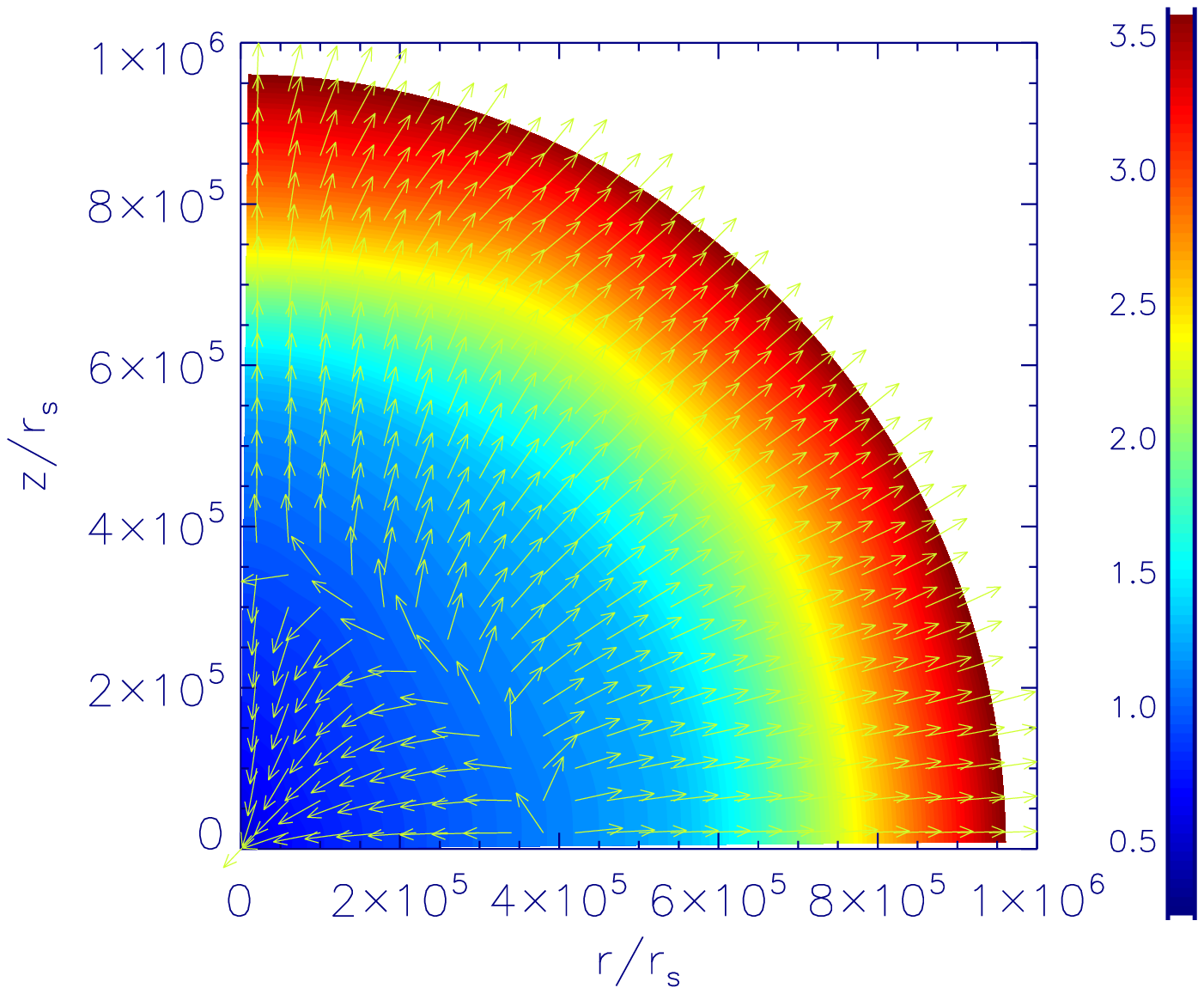}
\hspace*{0.5cm} \caption{Snapshot of the accretion flow properties
at $t=6\times 10^5$ year for model 1a. At this snapshot time, the wind is
present. Left: logarithm density (colour) overplotted by the poloidal
velocity vector (arrows). Middle: logarithm temperature (colour)
overplotted by the direction of the velocity vector. Right: the
ratio of thermal energy to the gravitational energy. It is clear, in
the wind region, the thermal energy is bigger than the gravitational
energy. \label{Fig:snapshota1}}
\end{center}
\end{figure*}

Figure \ref{Fig:snapshota1} shows snapshot of properties of accretion flow (for model 1a) at $t=6\times 10^5$ year when
wind is present. Wind forms in the region beyond the Compton and
Bondi radius $r>4\times 10^5 r_s$. In literatures, the widely
studied wind driven mechanisms include radiation pressure (e.g.,
Murray et al. 1995; Murray \& Chiang 1997; Proga 2007), magnetic driven (e.g.,
Blandford \& Payne 1982; Emmering et al. 1992; Romanova et al. 1997;
Bottorff et al. 2000; Cao 2011; Li \& Begelman 2014), and thermal
driven (e.g., Begelman et al. 1983; Chelouche \& Netzer 2005). In
this paper, the accretion flow is irradiated by a hot accretion flow
with low mass accretion rate; Therefore, the radiation pressure due
to Thomson scattering is not important. The only mechanism for
driving wind here is thermally driven. The right panel of Figure 1
shows the ratio between the thermal energy to the gravitational
energy of the flow. It is clear that the wind thermal energy is
bigger than the gravitational energy. The wind is thermally driven. We note that because the thermal energy of wind is bigger than its gravitational energy, the Bernoulli parameter (defined as the sum of gravitational energy, thermal energy and kinetic energy) of wind is positive. Wind can escape to infinity.

There are also some papers studying wind from a thin disk irradiated
by X-ray (e.g., Begelman et al. 1983; Woods et al. 1996; Chelouche
\& Netzer 2005; Higginbottom et al. 2017; Nomura \& Ohsuga 2017). It is found that beyond
the Compton radius, wind can be thermally driven above a thin disk.

To quantitatively study the properties of the inflow/outflow
component, we calculate the radial dependence of mass inflow,
outflow, and net rates as follows: (1) inflow rate
\begin{equation}
\dot {M}_{\rm in} (r)=4\pi r^2 \int_{\rm 0^\circ}^{\rm 90^\circ}
\rho \min (v_r, 0) \sin\theta d\theta
\end{equation}
(2) outflow rate
\begin{equation}
\dot {M}_{\rm out} (r)=4\pi r^2 \int_{\rm 0^\circ}^{\rm 90^\circ}
\rho \max (v_r, 0) \sin\theta d\theta
\end{equation}
and (3) net rate
\begin{equation}
\dot {M}_{\rm net} (r)=4\pi r^2 \int_{\rm 0^\circ}^{\rm 90^\circ}
\rho v_r \sin\theta d\theta
\end{equation}
We also calculate the kinetic and thermal energy carried by the wind
as follows:
\begin{equation}
P_{\rm k} (r)=2\pi r^2 \int_{\rm 0^\circ}^{\rm 90^\circ} \rho
\max(v_r^3,0) \sin\theta d\theta
\end{equation}
\begin{equation}
P_{\rm th} (r)=4\pi r^2 \int_{\rm 0^\circ}^{\rm 90^\circ} e
\max(v_r,0) \sin\theta d\theta
\end{equation}
The power of wind is defined as:
\begin{equation}
P_W=P_{\rm k}+P_{\rm th}
\label{power}
\end{equation}

\begin{figure}
\begin{center}
\includegraphics[scale=0.45]{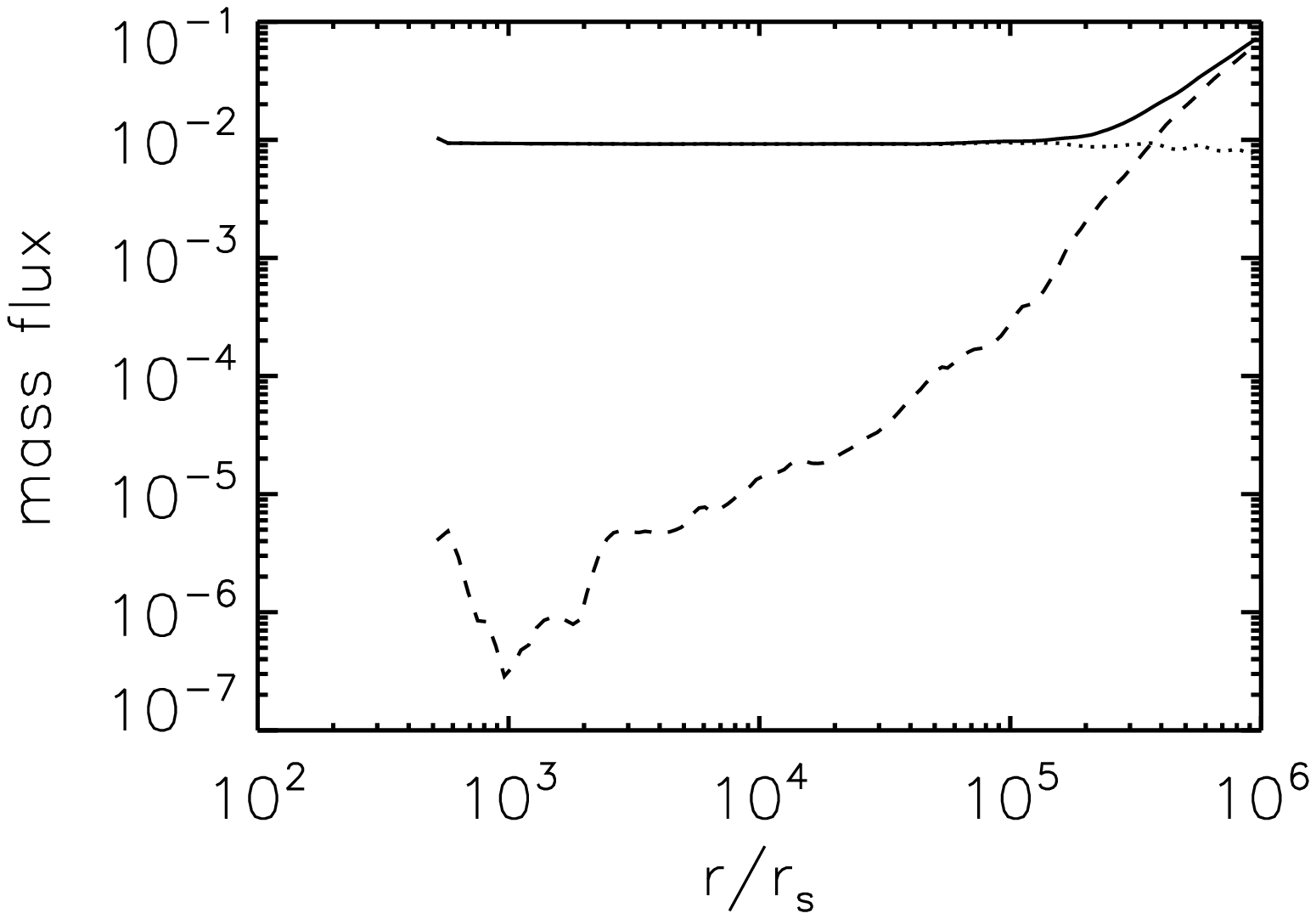}\hspace*{0.5cm}
\caption{The radial profiles of time-averaged (from $t=1.5 \times
10^5$ to $1.15 \times 10^6$ year) and angle integrated mass inflow
rate (solid line), outflow rate (dashed line) and the net rate
(dotted line) in model 1a. The mass fluxes are expressed in unit of the
Eddington accretion rate.\label{Fig:accretionrateA1}}
\end{center}
\end{figure}

\begin{figure}
\begin{center}
\includegraphics[scale=0.45]{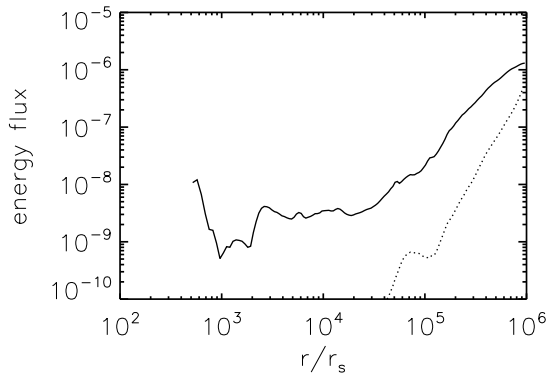}\hspace*{0.5cm}
\caption{The radial profiles of time-averaged (from $t=1.5 \times
10^5$ to $1.15 \times 10^6$ year) and angle integrated thermal (solid line) and kinetic (dotted line) powers carried by wind in model 1a. The
powers are expressed in unit of the Eddington Luminosity.\label{Fig:powerA1}}
\end{center}
\end{figure}

\begin{figure}
\begin{center}
\includegraphics[scale=0.45]{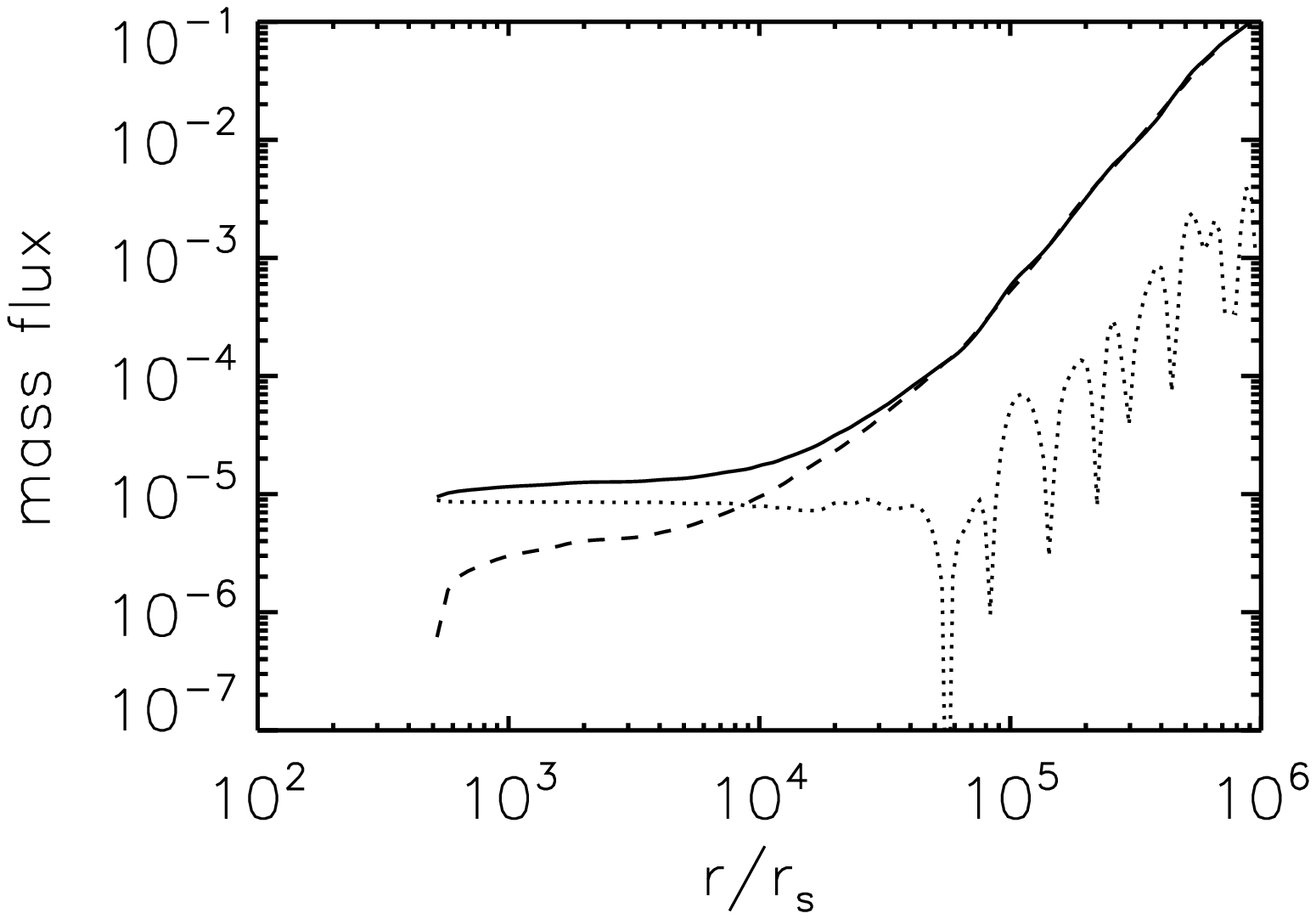}\hspace*{0.5cm}
\caption{The radial profiles of time-averaged (from $t=2 \times
10^5$ to $8.5 \times 10^5$ year) and angle integrated mass inflow
rate (solid line), outflow rate (dashed line) and the net rate
(dotted line) in model 4a. The mass fluxes are expressed in unit of the
Eddington accretion rate.\label{Fig:accretionrateA4}}
\end{center}
\end{figure}

\begin{figure}
\begin{center}
\includegraphics[scale=0.45]{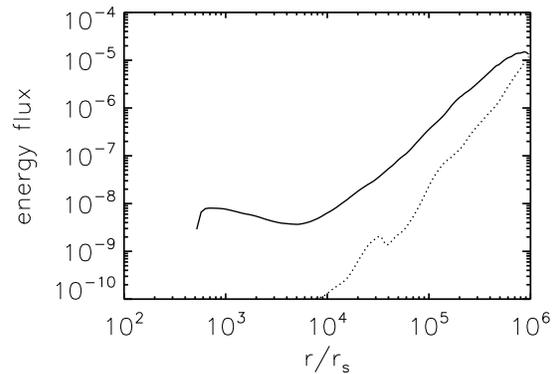}\hspace*{0.5cm}
\caption{The radial profiles of time-averaged (from $t=2 \times
10^5$ to $8.5 \times 10^5$ year) and angle integrated thermal power
(solid line) and kinetic power (dotted line) carried by wind in model \emph{}4a. The powers are expressed in unit of the Eddington
Luminosity.\label{Fig:powerA4}}
\end{center}
\end{figure}

Figure \ref{Fig:accretionrateA1} shows the radial profiles of time-averaged (from $t=1.5
\times 10^5$ to $1.15 \times 10^6$ year) and angle integrated mass
inflow rate (solid line), outflow rate (dashed line) and the net
rate (dotted line) in model 1a. The mass fluxes are expressed in unit
of the Eddington accretion rate. It is clear the wind
becomes important only beyond $3\times 10^5 r_s$. Beyond this
radius, the wind mass flux exceeds the net accretion rate. Inside
$3\times 10^5 r_s$, wind is very weak and almost all of the gas
falls onto the central black hole. This is consistent with that
shown in Figure 1.

Figure \ref{Fig:powerA1} shows the radial profiles of time-averaged (from $t=1.5
\times 10^5$ to $1.15 \times 10^6$ year) and angle integrated
thermal (solid line) and kinetic (dotted line) powers carried
by wind in model 1a. The powers are expressed in unit of the
Eddington Luminosity. It is clear that in the whole region, the
thermal energy carried by wind is larger than the kinetic energy. It
is another evidence that wind is thermally driven. Because wind thermally expands, it can be treated as an expanding `atmosphere' , similar to stellar atmospheres. We define the
wind production efficiency as
\begin{equation}
\varepsilon=P_W/\dot{M}_{\rm BH} c^2
\label{efficiency}
\end{equation}
$\dot{M}_{\rm BH}$ is calculated by using Equation (\ref{AccretionrateBH}). In model 1a,
we find that at the outer boundary $\varepsilon=10^{-4}$.

We study the effects of changing the outer boundary gas temperature by carrying out models 2a, 3a and 4a. Generally, we find that with the increase of gas temperature, the luminosity of the central AGN
decreases. The reason is easy to be understood. With the increase of
temperature, the Bondi radius decreases; The mass bound to the black
hole is decreased which results in a decrease of mass accretion rate
and luminosity.

We take model 4a as an example to illustrate the effects of changing
boundary temperature. In model 4a, the outer boundary
gas temperature is 5 times of that in model 1a. Figure \ref{Fig:accretionrateA4} shows the
radial profiles of time-averaged (from $t=2 \times 10^5$ to $8.5
\times 10^5$ year) and angle integrated mass fluxes in model 4a. The mass accretion rate
onto the black hole in model 4a is 3 orders of magnitude smaller than
that in model 1a. Figure \ref{Fig:powerA4} shows the radial profiles of time-averaged
(from $t=2 \times 10^5$ to $8.5 \times 10^5$ year) and angle
integrated thermal (solid line) and kinetic (dotted
line) powers carried by wind in model 4a. Comparing figures \ref{Fig:powerA4} and \ref{Fig:powerA1}, we find that at the outer boundary, the thermal energy flux in model 4a is much higher than that in model 1a. The reason is as follows. Because thermal energy flux is proportional to mass flux of wind times wind temperature. The mass flux of wind at the outer boundary in model 1a is comparable to that in model 4a. However, temperature of wind in model 4a is much higher than that in model 1a. Therefore, wind energy flux in model 4a is higher than that in model 1a. The thermal energy carried by wind in model 4a
is larger than kinetic energy. We find that at the outer boundary
$\varepsilon=1.18$ in model 4a.

We study the effects of changing outer boundary density by carrying out models 2ad, 3ad and 4ad. We summarize our results in Table 1. We find that with the increase of density, the mass accretion rate onto the black hole can be significantly increased. The reasons are as follows. As we mentioned above, the temperature of the accretion flow is found to be higher than $10^7$K. The dominating heating and cooling processes are Compton heating and  bremsstrahlung cooling. The Compton heating (Equation (\ref{Compton})) is $\propto \rho$, the bremsstrahlung cooling (Equation (\ref{brem})) is $\propto \rho^2$. With the increase of density, the cooling increases faster than heating. Therefore, the temperature will decrease with increasing density. With the decrease of temperature (or gas pressure gradient force in radial direction), the infall velocity of gas increases. Therefore, with the increase of infall velocity and density, the mass accretion rate onto the black hole increases. Due to the significant increase of mass accretion rate, the wind production efficiency decreases significantly. We find that when the
environment density at pc-scale is higher than $4\times10^{-22}
\text{g } {\rm cm}^{-3}$, the luminosity of the central AGN exceeds 2$\% L_{Edd}$. In this case, the spectrum of black
hole has a chance to transit from hard to soft state. We do not discuss this case in which the luminosity of the central AGN exceeds 2$\% L_{Edd}$, because we focus on accretion flow irradiated by a LLAGN.

We take models 3a and 3ad as examples to discuss the effects of changing outer boundary density (see Table 1 for the results). The only difference between models 3a and 3ad is the outer boundary density. The outer boundary density in model 3ad is 3 times higher than that in model 3a. We can see that the mass accretion rate in model 3ad is more than 2 orders of magnitude higher than that in model 3a. The wind mass flux in model 3ad is 8 times of that in model 3a. The increase of wind mass flux is due to the increase of density in the wind region. With the increase of wind mass flux, the power of wind increases. We note that the power of wind is increased by a factor smaller than 8. Because $P_{\rm w} \propto \dot {M}_{\rm w} \times T$. Due to the decrease of wind temperature in model 3ad, the increase of wind power is smaller than $8$. Due to the high accretion rate in model 3ad, the production efficiency of wind is much smaller than that in model 3a.

We also study the effects of Compton temperature by carrying out models 1aTx, 2aTx and 3aTx. The results are summarized in Table 1. We find that when the temperature of accretion flow is low, the effects of increasing Compton temperature are obvious (see the comparision of results in models 1aTx and 1a presented in table 1). The effects of increasing Compton temperature become less important with the increase of accretion flow temperature (see the comparision of results in models 3aTx and 3a presented in table 1). The reason is as follows. According to equations (4), (5) and (6), the Compton heating in a unit volume is proportional to $L(T_X-T)n$. Therefore, for gas of unit mass, the Compton heating is proportional to $L(T_X-T$). X-ray luminosity is proportional to mass accretion rate. Therefore, Compton heating is proportional to $\dot{M}(T_X-T)$. Although, the increase of $(T_X-T)$ from model 1a to 1aTx is comparable to the increase of $(T_X-T)$ from model 3a to 3aTx. The mass accretion rates in models 1a and 1aTx are about 2 orders of magnitude higher than those in models 3a and 3aTx (The value of $\dot M$ is presented in Table 1).  Therefore, the effects of increasing Compton temperature are more obvious in low temperature accretion flow case.  We take models 1aTx and 1a as examples to discuss the effects of increasing Compton temperature. We find that accretion rate decreases by a factor of $\sim 8$ in model 1aTx compared to that in model 1a. The reason is as follows. Higher Compton temperature results in stronger Compton heating and higher gas temperature. When accretion flow temperature becomes higher, the gas infall velocity decreases due to stronger gas pressure gradient force. The accretion rate decreases with decreasing infall velocity. Also, higher temperature can also result in larger wind power, because higher temperature gas has larger specific energy.

\begin{figure*}
\begin{center}
\includegraphics[scale=0.3]{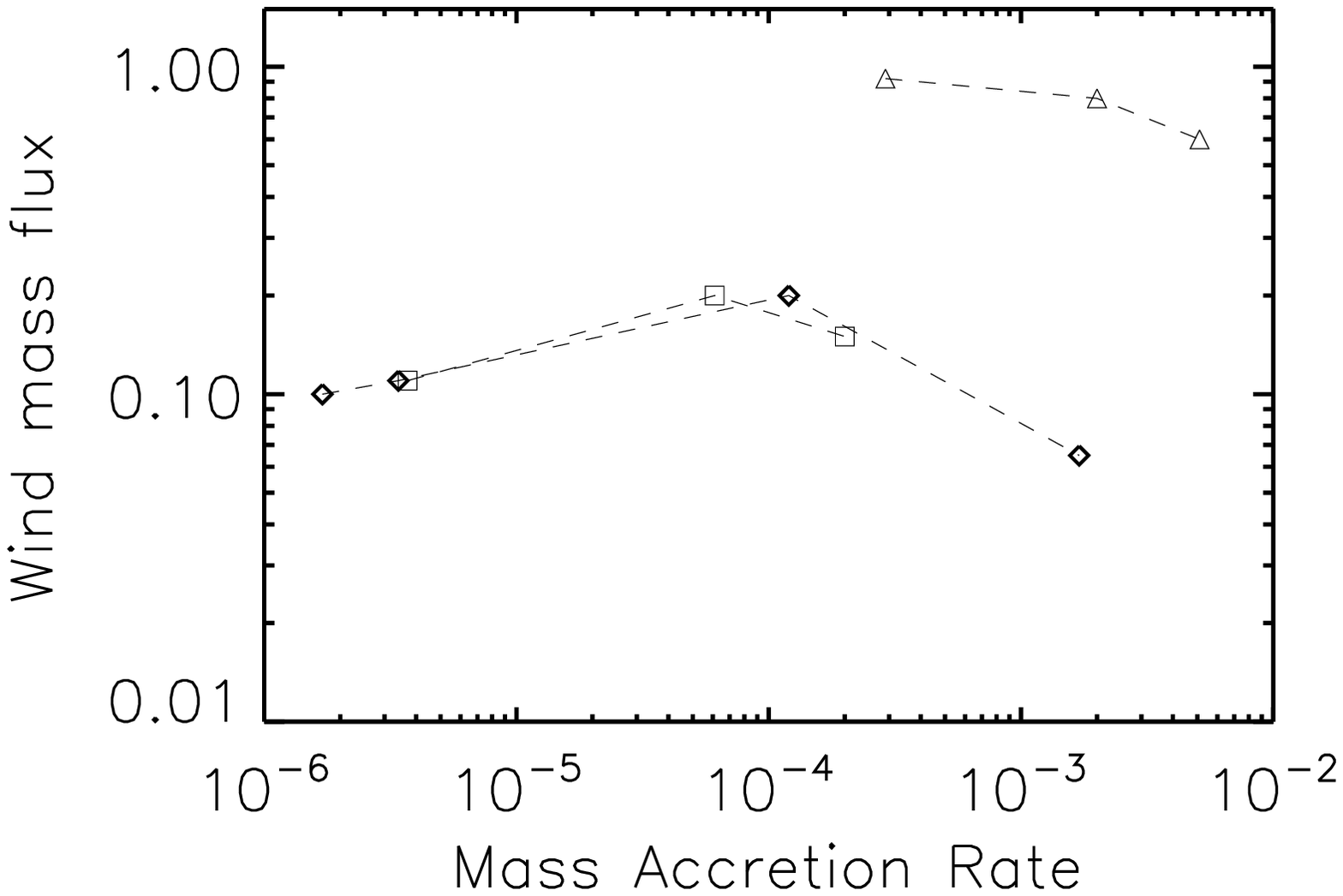}\hspace*{0.3cm}
\includegraphics[scale=0.3]{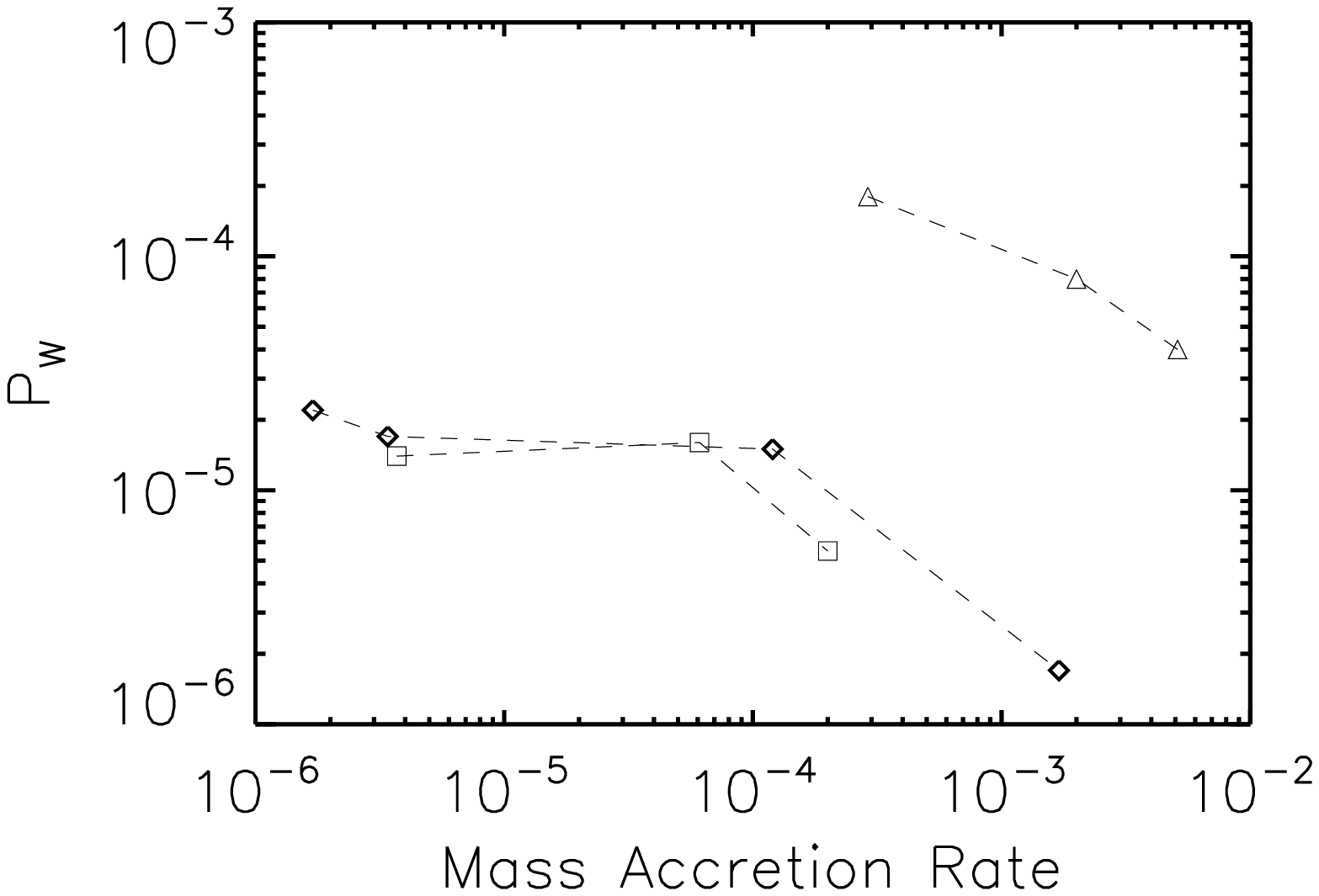}\hspace*{0.3cm}
\includegraphics[scale=0.3]{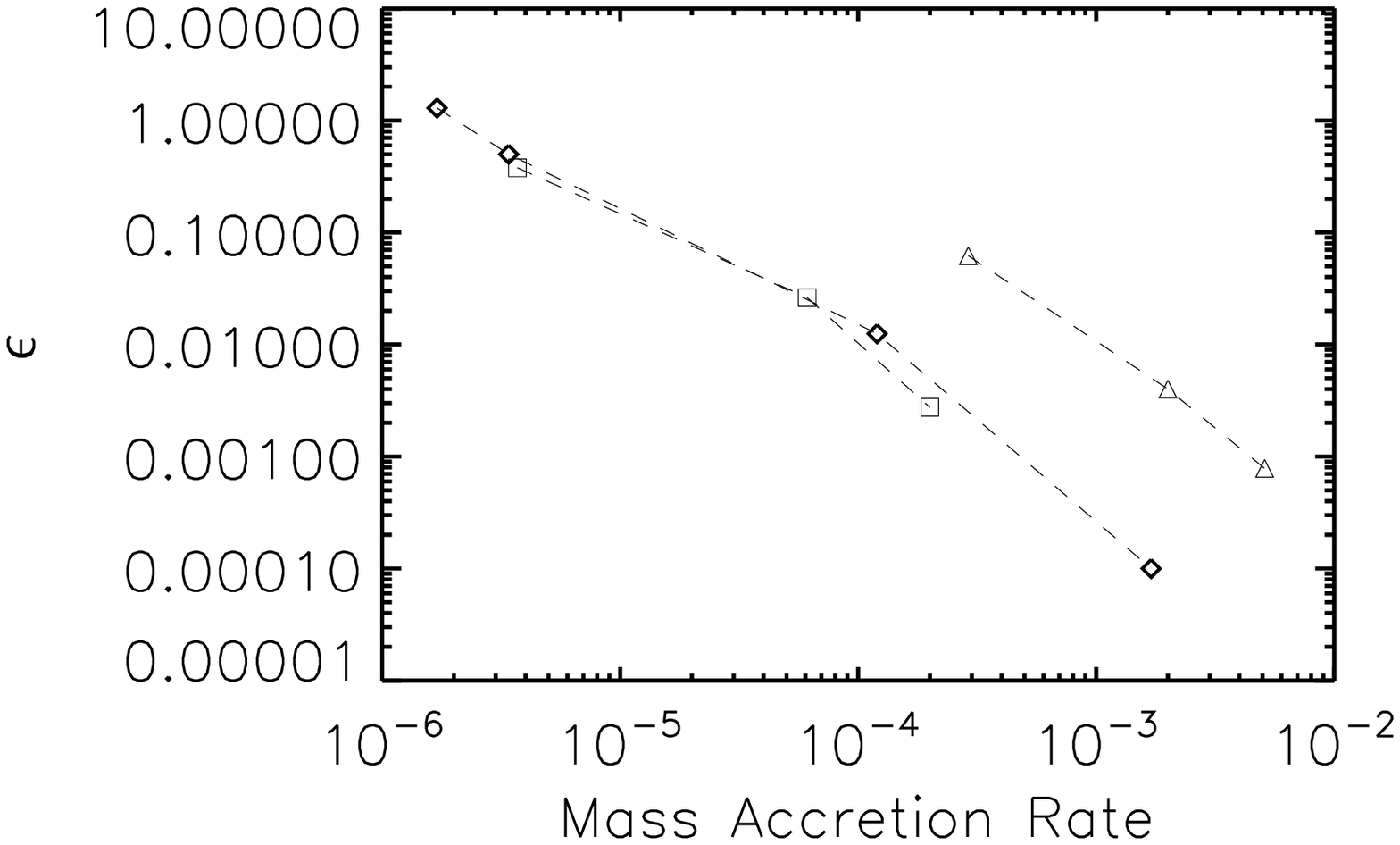}
\hspace*{0.3cm} \caption{Summary of the results presented in table 1. In this figure, diamonds represent results from models 1a-4a. Triangles represent results from models 2ad-4ad. Squares represent results from models 1aTx-3aTx. The horizontal axis in each panel is accretion rate onto the black hole in unit of Eddington accretion rate. Left: the wind mass flux (in unit of Eddington accretion rate) measured at the outer boundary of the simulation as a function of mass accretion rate. Middle: power of wind (in unit of Eddington luminosity, see Equation (\ref{power})) measured at the outer boundary of the simulation as a function of mass accretion rate onto the black hole. Right: wind production efficiency (see Equation (\ref{efficiency})) measured at the outer boundary of the simulation as a function of mass accretion rate onto the black hole. \label{Fig:summary}}
\end{center}
\end{figure*}

In order to summarize the results discussed above, we plot Figure \ref{Fig:summary}. In this figure, diamonds represent results from models 1a-4a. Triangles represent results from models 2ad-4ad. Squares represent results from models 1aTx-3aTx. The mass flux of wind is in the range $10^{-2}<\dot {M}_W/\dot{M}_{\rm Edd}<1$. The power of wind is in the range $10^{-6}<\dot P_W/\dot{L}_{\rm Edd}<10^{-3}$. From this figure, we find that the changing of Compton temperature has small effects on results. As introduced above, with the increase of gas density at the outer boundary, the properties of wind change significantly. For example, with the increase of gas density at the outer boundary, wind mass flux and power increase significantly. Correspondingly, the production efficiency of wind increases significantly. Note that the power of wind does not strongly correlate with mass accretion rate, therefore, the wind production efficiency is predominately determined by the black hole accretion rate. Therefore, generally, the production efficiency of wind decreases with increasing mass accretion rate. We note that the production efficiency of wind can be larger than $1$, when $\dot {M}_W/\dot{M}_{\rm Edd} \sim 10^{-6}$. Yuan et al. (2015) studied wind generated in hot accretion flow with larger angular momentum in the region very close to the black hole. They find that wind can be generated by the combination of centrifugal and magnetic pressure gradient forces. In that work, it is found the wind production efficiency is $\sim 1/1000$.

\section{Summary and discussion}
In this paper, we perform two-dimensional hydrodynamical simulations to study slowly rotating accretion flow at parsec and sub-parsec scale irradiated by a LLAGN. We take into account the Compton heating and photoionization heating by the central X-ray. The bremsstrahlung cooling, recombination and line cooling are also taken into account. The temperature of the accretion flow is found to be above $10^7$K. Therefore, the photoionization heating, recombination and line cooling are negligibly small. The dominate heating and cooling precesses are Compton heating and bremsstrahlung cooling. We find that due to the Compton heating by the central X-ray, the accretion flow temperature can be above the virial temperature. Wind can be thermally driven. The mass flux of wind is in the range $10^{-2}<\dot {M}_W/\dot{M}_{\rm Edd}<1$. The power of wind is in the range $10^{-6}<\dot P_W/\dot{L}_{\rm Edd}<10^{-3}$. We find the wind production efficiency decreases with increasing mass accretion rate. The production efficiency of wind is in the range $10^{-4} < \epsilon \leq 1.18$.

In the large scale AGN feedback simulations (e.g., Ciotti et al. 2010; Gaspari et al. 2012), ``mechanical feedback" by AGN wind is usually involved to heat the intercluster medium to prevent rapid cool of the gas (i.e., the cooling flow problem). It is found that to be consistent with observations, the production efficiency should be larger than $10^{-4}$. Therefore, the thermally driven wind may play a role in solving the rapid cooling problem of intercluster medium when the AGN is in hot accretion mode.

In this paper, we study slowly rotating accretion flow. In future, it is necessary to study accretion flow with high angular momentum. If the angular momentum of accretion flow is high, a rotating disk can form. Viscosity is needed to transfer angular momentum. There are several big differences between slowly rotating flow and high angular momentum accretion disk. First, in high angular momentum accretion disk, viscous heating can increase the gas temperature; Second, in the high angular momentum accretion flow, gas can fall onto the central black hole on a viscous timescale. Viscous timescale is much longer than the nearly free fall timescale of slowly rotation flow. Therefore, there is a longer time lag between the changing of mass accretion rate at large scale and the response in generation of X-ray photons close to the black hole. Third, in the high angular momentum accretion flow, centrifugal force helps to generate wind.

In accretion flows, ordered, large-scale open magnetic field may exist (e.g. Blandford \& Payne 1982; Lovelace et al. 1994; Cao 2011; Penna et al. 2013; Bai \& Stone 2013; Li \& Begelman 2014). In addition to thermally driven wind, magneto-centrifugal wind may also be present (Blandford \& payne 1982; Cao 2011; Li \& Begelman 2014). Therefore, it is necessary to study the accretion flow at parsec scale with magnetic field in future.

\section*{Acknowledgments}
We thank the referee for his/her valuable comments which help to improve this paper significantly. This work is supported in part by the National Program on Key Research and Development Project
of China (Grant No. 2016YFA0400704),  the Natural Science Foundation of China (grants
11573051, 11633006, 11773053 and 11661161012), the Natural Science
Foundation of Shanghai (grant 16ZR1442200), and the Key
Research Program of Frontier Sciences of CAS (No. QYZDJSSW-
SYS008).  This work made use of the High Performance Computing Resource in the Core
Facility for Advanced Research Computing at Shanghai Astronomical
Observatory.


\begin{thebibliography}{99}
\bibitem[\protect\citeauthoryear{Abramowicz et al.}{1991}]{Abramowicz et al. 1991} Abramowicz M. A., Chen X., Kato S., Lasota J. P., Regev O., 1995, ApJL, 438, L37
\bibitem[\protect\citeauthoryear{Antonucci}{2012}]{Antonucci 2012} Antonucci R., 2012, A\&AT, 27, 557
\bibitem[\protect\citeauthoryear{Bai \& Stone}{2013}]{Bai and Stone 2013} Bai X. N., Stone J. M., 2013, ApJ, 767, 30
%\bibitem[\protect\citeauthoryear{Blaes et al.}{2006}]{blaes et al. 2006} Blaes O. M., Arras P., Fragile P. C., 2006, MNRAS, 369, 1235
\bibitem[\protect\citeauthoryear{Beckwith et al.}{2008}]{Beckwith et al. 2008}Beckwith K., Hawley J. F., Krolik J. H., 2008, ApJ, 678, 1180
\bibitem[\protect\citeauthoryear{Begelman et al.}{1983}]{Begelman et al. 1983} Begelman M. C., McKee C. F., Shields G. A., 1983, ApJ, 271, 70
%\bibitem[\protect\citeauthoryear{Begelman}{2012}]{Begelman 2012} Begelman M. C., 2012, MNRAS, 420, 2912
\bibitem[\protect\citeauthoryear{Belloni}{2010}]{Belloni 2010} Belloni T. M., 2010, in Belloni T., ed., Lecture Notes in Physics, Vol. 794, The Jet Paradigm - From Microquasars to Quasars. Springer-Verlag, Berlin, p. 53
%\bibitem[\protect\citeauthoryear{Blackman et al. 2008}{2008}]{Blackman et al. 2008} Blackman E. G., Penna R. F., Varniere P., 2008,  NewA, 13, 244
\bibitem[\protect\citeauthoryear{Blandford \& Payne}{1982}]{Blandford and Payne 1982} Blandford R. D., Payne D. G., 1982, MNRAS, 199, 883
%\bibitem[\protect\citeauthoryear{Booth \& Schaye}{2009}]{Booth and Schaye 2009} Booth C. M., Schaye J., 2009, MNRAS, 398, 53
\bibitem[\protect\citeauthoryear{Bottorff et al. 2000}{2000}]{Bottorff et al. 2000} Bottorff M. C., Korista K. T., Shlosman I., 2000, ApJ, 537, 134
\bibitem[\protect\citeauthoryear{Bu et al. 2013}{2016}]{Bu et al. 2013} Bu D. F., Yuan F., Wu M. C., Cuadra J. 2013, MNRAS, 434, 1692
\bibitem[\protect\citeauthoryear{Bu et al. 2016a}{2016}]{Bu et al. 2016a} Bu D. F., Yuan F., Gan Z. M., Yang X. H. 2016a, ApJ, 818, 83
\bibitem[\protect\citeauthoryear{Bu et al. 2016b}{2016}]{Bu et al. 2016b} Bu D. F., Yuan F., Gan Z. M., Yang X. H. 2016b, ApJ, 823, 90
\bibitem[\protect\citeauthoryear{Cao 2011}{2011}]{Cao 2011} Cao X. W., 2011, ApJ, 737, 94
%\bibitem[\protect\citeauthoryear{Clarke 1996}{1996}]{Clarke 1996} Clarke D., 1996, ApJ, 457, 291
\bibitem[\protect\citeauthoryear{Chelouche \& Netzer 2005}{2005}]{Chelouche and Netzer 2005} Chelouche D., Netzer H. 2005, ApJ, 625, 95
\bibitem[\protect\citeauthoryear{Cheung et al. 2016}{2016}]{Cheung et al. 2016} Cheung E., Bundy K., Cappellari M., et al. 2016, Nature, 533, 504
\bibitem[\protect\citeauthoryear{Ciotti \& Ostriker 2007}{2007}]{Ciotti and Ostriker 2007} Ciotti L., Ostriker J. P., 2007, ApJ, 665, 1038
%\bibitem[\protect\citeauthoryear{Ciotti et al. 2017}{2017}]{Ciotti et al. 2017} Ciotti L., Pellegrini S., Negri A., Ostriker J. P., 2017, ApJ, 835, 15
\bibitem[\protect\citeauthoryear{Ciotti et al. 2010}{2010}]{Ciotti et al. 2010} Ciotti L., Ostriker J. P., Proga D., 2010, ApJ, 717, 708
%\bibitem[\protect\citeauthoryear{Cowie et al. 1978}{1978}]{Cowie et al. 1978} Cowie L. L., Ostriker J. P., Stark A. A., 1978, ApJ, 226, 1041
\bibitem[\protect\citeauthoryear{Crenshaw \& Kraemer 2012}{2012}] {Crenshaw and Kraemer 2012} Crenshaw D. M., Kraemer S. B., 2012, ApJ, 753, 75
\bibitem[\protect\citeauthoryear{De Villiers et al. 2003}{2003}]{De Villiers et al. 2003} De Villiers J. P., Hawley J. F., Krolik J. H., 2003, ApJ, 599, 1238
\bibitem[\protect\citeauthoryear{Di Matteo al. 2005}{2005}]{Di Matteo et al. 2005} Di Matteo T., Springel V., Hernquist L., 2005, Natur, 433, 604
\bibitem[\protect\citeauthoryear{Done 2014}{2014}]{Done 2014} Done C., 2014, in Ishida M., Petre R., Mitsuda K., eds, Suzaku-MAXI 2014: Expanding the Frontiers of the X-ray Universe. Ehime University, Matsuyama, Ehime Prefecture, Japan, p. 300
\bibitem[\protect\citeauthoryear{Esin et al. 1997}{1997}]{Esin et al. 1997} Esin A. A., McClintock J. E., Narayan R., 1997, ApJ, 489, 865
\bibitem[\protect\citeauthoryear{Emmering et al. 1992}{1992}]{Emmering et al. 1992} Emmering R. T., Blandford R. D., Sholosman I., 1992, ApJ, 385, 460
\bibitem[\protect\citeauthoryear{Fender et al. 2004}{2004}]{Fender et al. 2004} Fender R. P., Belloni T. M., Gallo E, 2004, MNRAS, 355, 1105
\bibitem[\protect\citeauthoryear{Ferrarese \& Merritt 2000}{2000}]{Ferrarese and Merritt 2000} Ferrarese L., Merritt D., 2000, ApJ, 539, 9
%\bibitem[\protect\citeauthoryear{Fishbone \& Moncrief 1976}{1976}]{Fishbone and Moncrief 1976} Fishbone L. G., Moncrief V., 1976, ApJ, 207, 962
\bibitem[\protect\citeauthoryear{Gan et al. 2014}{2014}]{Gan et al. 2014} Gan Z. M., Yuan F., Ostriker J. P., Ciotti L., Novak G. S., 2014, ApJ, 789, 150
\bibitem[\protect\citeauthoryear{Gaspari et al. 2012}{2012}]{Gaspari et al. 2012} Gaspari M., Brighenti F., Temi P., 2012, MNRAS, 424, 190
\bibitem[\protect\citeauthoryear{Gebhardt et al. 2000}{2000}]{Gebhardt et al. 2000} Gebhardt K., Bender R., Bower G., 2000, ApJ, 539, L13
%\bibitem[\protect\citeauthoryear{Gu 2015}{2015}]{Gu 2015} Gu W. M., 2015, ApJ, 799, 71
%\bibitem[\protect\citeauthoryear{Guan et al. 2009}{2009}]{Guan et al. 2009} Guan X., Gammie C. F., Simon J. B., Johnson B. M., 2009, ApJ, 694, 1010
\bibitem[\protect\citeauthoryear{Hawley et al. 2001}{2001}]{Hawley et al. 2001} Hawley J. F., Balbus S. A., Stone J. M., 2001, ApJ, 554, L49
%\bibitem[\protect\citeauthoryear{Hawley \& Balbus 2002}{2002}]{Hawley and Balbus 2002}  Hawley J. F., Balbus S. A., 2002, ApJ, 573, 738
\bibitem[\protect\citeauthoryear{Hayes et al. 2006}{2006}]{Hayes et al. 2006} Hayes J. C., Norman M. L., Fiedler R. A., Bordner J. O., Li P. S., 2006, ApJ, 165, 188
\bibitem[\protect\citeauthoryear{Higginbottom et al. 2017}{2017}]{Higginbottom et al. 2017} Higginbottom N., Proga D., knigge C., Long K. S., 2017, ApJ, 836, 42
\bibitem[\protect\citeauthoryear{Ho 2008}{2008}]{Ho 2008} Ho L. C., 2008, ARA\&A, 46, 475
\bibitem[\protect\citeauthoryear{Homan et al. 2016}{2016}]{Homan et al. 2016} Homan J., Neilson J., Allen J. L., et al. 2016, ApJ, 830, L5
\bibitem[\protect\citeauthoryear{Inayosi et al. 2017}{2017}]{Inayoshi et al. 2017} Inayoshi K., Ostriker J. P., Haiman Z., Kuiper R., 2017 (arXiv: 1709.07452)
\bibitem[\protect\citeauthoryear{Igumenshchev \& Abramowicz 1999}{1999}]{Igumenshchev and Abramowicz 1999} Igumenshchev I. V., Abramowicz M. A., 1999, MNRAS, 303, 309
\bibitem[\protect\citeauthoryear{Igumenshchev \& Abramowicz 2000}{2000}]{Igumenshchev and Abramowicz 2000} Igumenshchev I. V., Abramowicz M. A., 2000, ApJS, 130, 463
\bibitem[\protect\citeauthoryear{Kato et al. 1998}{1998}]{Kato et al. 1998} Kato S., Fukue J., Mineshige S., 1998, Black Hole Accretion Disks. Kyoto Univ. Press, Kyoto
\bibitem[\protect\citeauthoryear{Kormendy \& Bender 2009}{2009}]{Kormendy and Bender 2009} Kormendy J., Bender R., 2009, ApJ, 691, L142
\bibitem[\protect\citeauthoryear{Kurosawa \& Proga 2009}{2009}]{Kurosawa and Proga 2009} Kurosawa R., Proga D., 2009, MNRAS, 397, 1791
\bibitem[\protect\citeauthoryear{Lovelace et al. 1994}{1994}]{Lovelace et al. 1994} Lovelace R. V. E., Romanova M. M., Newman W. I., 1994, ApJ, 437, 136
\bibitem[\protect\citeauthoryear{Li et al. 2013}{2013}]{Li et al. 2013} Li J., Ostriker J., Sunyaev R., 2013, ApJ, 767, 105
\bibitem[\protect\citeauthoryear{Li \& Begelman 2014}{2014}]{Li and Begelman 2014} Li S. L., Begelman M. C., 2014, ApJ, 786, 6
\bibitem[\protect\citeauthoryear{Liu et al. 2013}{2013}]{Liu et al. 2013} Liu C., Yuan F., Ostriker J., Gan Z., Yang X., 2013, MNRAS, 434, 1721
\bibitem[\protect\citeauthoryear{Machida et al. 2001}{2001}]{Machida et al. 2001} Machida M., Matsumoto R., Mineshige S., 2001, PASJ, 53, L1
\bibitem[\protect\citeauthoryear{Magorrian et al. 2001}{2001}]{Magorrian et al. 2001} Magorrian J., Tremaine S., Richstone D. et al. 1998, AJ, 115, 2285
\bibitem[\protect\citeauthoryear{McKinney et al. 2012}{2012}]{McKinney et al. 2012} McKinney J., Tchekhovskoy A., Blandford R., 2012, MNRAS, 423, 3083
\bibitem[\protect\citeauthoryear{Moller \& Sadowski 2015}{2015}]{Moller and Sadowski 2015} Moller A., Sadowski A., 2015 (arXiv:1509.06644)
\bibitem[\protect\citeauthoryear{Mo\'{s}cibrodzka et al. 2014}{2014}]{Moscibrodzka et al. 2014} Mo\'{s}cibrodzka M., Falcke H., Shiokawa H., Gammie C. F., 2014, A\&A, 570, A7
\bibitem[\protect\citeauthoryear{Murray et al. 1995}{1995}]{Murray et al. 1995} Murray N., Chiang J., Grossman S. A., Voit G. M., 1995, ApJ, 451, 498
\bibitem[\protect\citeauthoryear{Murray \& Chiang 1997}{1997}]{Murray and Chiang 1997} Murray N., Chiang J., 1997, ApJ, 474, 91
\bibitem[\protect\citeauthoryear{Narayan \& Yi 1994}{1994}]{Narayan and Yi 1994} Narayan R., Yi I., 1994, ApJ, 428, L13
\bibitem[\protect\citeauthoryear{Narayan \& Yi 1995}{1995}]{Narayan and Yi 1995} Narayan R., Yi I., 1995, ApJ, 452, 710
\bibitem[\protect\citeauthoryear{Narayan et al. 1998}{1998}]{Narayan et al. 1998} Narayan R., Mahadevan R., Quataert E., 1998, in Abramowicz M. A., Bjornsson G., Pringle J. E., eds, Theory of Black Hole Accretion Discs. Cambridge Univ. Press, Cambridge, p. 148
\bibitem[\protect\citeauthoryear{Narayan \& McClintock 2008}{2008}]{Narayan and McClintock 2008} Narayan R., McClintock J. E., 2008, New Astron. Rev., 51, 733
\bibitem[\protect\citeauthoryear{Narayan et al. 2012}{2012}]{Narayan et al. 2012}Narayan R., Sadowski A., Penna R. F., Kulkarni A. K., 2012, MNRAS, 426, 3241
\bibitem[\protect\citeauthoryear{Nomura \& Ohsuga 2017}{2017}]{Nomura and Ohsuga 2017} Nomura M., Ohsuga K., 2017, MNRAS, 465, 2873
\bibitem[\protect\citeauthoryear{Ostriker et al. 2010}{2010}]{Ostriker et al. 2010} Ostriker J. P., Choi E., Ciotti L., Novak G. S., Proga D., 2010, ApJ, 722, 642
\bibitem[\protect\citeauthoryear{Pang et al. 2011}{2011}]{Pang et al. 2011} Pang B., Pen U.-L., Matzner C. D., Green S. R., Liebendorfer M., 2011, MNRAS, 415, 1228
\bibitem[\protect\citeauthoryear{Pen et al. 2003}{2003}]{Pen et al. 2003} Pen U. L., Matzener C. D., Wong S., 2003, ApJ, 596, L207
\bibitem[\protect\citeauthoryear{Penna et al. 2013}{2013}]{Penna et al. 2013} Penna R. F., Sadowski A., Kulkarni A. K., Narayan R., 2013, MNRAS, 428, 2255
%\bibitem[\protect\citeauthoryear{Power et al. 2011}{2011}]{Power et al. 2011} Power C., Nayakshin S., King A., 2011, MNRAS, 412, 269
%\bibitem[\protect\citeauthoryear{Proga et al. 2000}{2000}]{Proga et al. 2000} Proga D., Stone J. M., Kallman T. R., 2000, ApJ, 543, 686
\bibitem[\protect\citeauthoryear{Proga 2007}{2007}]{Proga 2007} Proga D., 2007, ApJ, 661, 693
\bibitem[\protect\citeauthoryear{Remillard \& McClintock 2006}{2006}]{Remillard and McClintock 2006} Remillard R. A., McClintock J. E., 2006, ARA\&A, 44, 49
\bibitem[\protect\citeauthoryear{Romanova et al. 1997}{1997}]{Romanova et al. 1997} Romanova M. M., Ustyugova G. V., Koldoba A. V., Chechetkin V. M., Lovelace R. V. E., 1997, ApJ, 482, 708
\bibitem[\protect\citeauthoryear{Sadowski et al. 2013}{2013}]{Sadowski et al. 2013} Sadowski A., Narayan R., Penna R., Zhu Y., 2013, MNRAS, 436, 3856
\bibitem[\protect\citeauthoryear{Sazonov et al. 2005}{2005}]{Sazonov et al. 2005} Sazonov S. Y., Ostriker J. P., Ciotti L., Sunyaev R. A., 2005, MNRAS, 358, 168
\bibitem[\protect\citeauthoryear{Stone et al. 1999}{1999}]{Stone et al. 1999} Stone J. M., Pringle J. E., Begelman M. C., 1999, MNRAS, 310, 1002
\bibitem[\protect\citeauthoryear{Tchekhovskoy et al. 2011}{2011}]{Tchekhovskoy et al. 2011} Tchekhovskoy, A., Narayan R., McKinney J. C., 2011, MNRAS, 418, L79
%\bibitem[\protect\citeauthoryear{Tchekhovskoy \& McKinney}{2012}]{Tchekhovskoy and McKinney 2012} Tchekhovskoy, A., McKinney J. C., 2012, MNRAS, 423, L55
\bibitem[\protect\citeauthoryear{Tombesi et al. 2010}{2010}]{Tombesi et al. 2010} Tombesi F., Sambruna J. N., Reeves J. N., et al. 2010, ApJ, 719, 700
\bibitem[\protect\citeauthoryear{Tombesi et al. 2014}{2014}]{Tombesi et al. 2014} Tombesi F., Tazaki F., Mushotzky R. F., et al. 2014, MNRAS, 443, 2154
\bibitem[\protect\citeauthoryear{Wang et al. 2013}{2013}]{Wang et al. 2013} Wang Q. D., Nowak M. A., Markoff S. B., et al., 2013, Sci, 341, 98
\bibitem[\protect\citeauthoryear{Woods et al. 1996}{1996}]{Woods et al. 1996} Woods D. T., Klein R. I., Castor J. I., McKee C. F., Bell J. B., 1996, ApJ, 461, 767
\bibitem[\protect\citeauthoryear{Wu et al. 2013}{2013}]{Wu et al. 2013} Wu Q. W., Cao X., Ho L. C., Wang D. X., 2013, ApJ, 700, 31
%\bibitem[\protect\citeauthoryear{Wu et al. 2016}{2016}]{Wu et al. 2016} Wu M. C., Xie F. G., Yuan Y. F., Gan Z. M., 2016, MNRAS, 459, 1543
%\bibitem[\protect\citeauthoryear{Xie et al. 2010}{2010}]{Xie et al. 2010} Xie F. G., Niedzwiecki A., Zdziarski A. A., Yuan F., 2010, MNRAS, 403, 170
%\bibitem[\protect\citeauthoryear{Yuan 2001}{2001}]{Yuan 2002} Yuan F., 2001, MNRAS, 324, 119
%\bibitem[\protect\citeauthoryear{Yuan et al. 2003}{2003}]{Yuan et al. 2003} Yuan F., Quataert E., Narayan R., 2003, ApJ, 598, 301
%\bibitem[\protect\citeauthoryear{Yuan 2003}{2003}]{Yuan 2003} Yuan F., 2003, ApJ, 594, L99
\bibitem[\protect\citeauthoryear{Xie \& Yuan 2012}{2012}]{Xie and Yuan 2012} Xie F., Yuan F., 2012, ApJ, 427, 1580
\bibitem[\protect\citeauthoryear{Xie et al. 2017}{2017}]{Xie et al. 2017} Xie F., Yuan F., Ho L. C., 2017, ApJ, 844, 42
\bibitem[\protect\citeauthoryear{Yuan et al. 2009}{2009}]{Yuan et al. 2009} Yuan F., Xie F., Ostriker J. P., 2009, ApJ, 691, 98
\bibitem[\protect\citeauthoryear{Yuan \& Li 2011}{2011}]{Yuan and Li 2011} Yuan F., Li M., 2011, ApJ, 737, 23
\bibitem[\protect\citeauthoryear{Yuan et al. 2012a}{2012}]{Yuan et al. 2012} Yuan F., Wu M. C., Bu D. F., 2012a, ApJ, 761, 129
\bibitem[\protect\citeauthoryear{Yuan et al. 2012b}{2012}]{Yuan et al. 2012} Yuan F., Bu D. F., Wu M. C., 2012b, ApJ, 761, 130
\bibitem[\protect\citeauthoryear{Yuan \& Narayan 2014}{2014}]{Yuan and Narayan 2014} Yuan F., Narayan R., 2014, ARA\&A, 52, 529
\bibitem[\protect\citeauthoryear{Yuan et al. 2015}{2015}]{Yuan et al. 2015} Yuan F., Gan Z. M., Narayan R., Sadowski A., Bu D. F., Bai X. N., 2015, ApJ, 804, 101
\bibitem[\protect\citeauthoryear{Zdziarski \& Gierlinski 2004}{2004}]{Zdziarski and Gierlinski 2004} Zdziarski A. A., Gierli\'{n}ski M., 2004, PThPS, 155, 99

\end{thebibliography}
\end{document}